\documentclass[twocolumn,showkeys,showpacs,preprintnumbers,prd,superscriptaddress,nofootinbib]{revtex4}
\usepackage{graphicx}
\usepackage{epsf}
\usepackage{bm}
\usepackage{amsmath}
\usepackage{amsfonts}
\usepackage{amssymb}
\usepackage{epstopdf}
\usepackage{natbib}
\usepackage{color}
\usepackage{verbatim}
\usepackage{caption}
\usepackage{multirow}
\usepackage{bm}
\usepackage{listings}
\usepackage{soul}
\usepackage{caption}
\usepackage{subcaption}
\usepackage{ragged2e}
\usepackage{float}
\usepackage{hyperref}

\begin{document}

\title{Investigating a Possible Variation of the Gravitational Constant Through Gas Mass Fraction Measurements and Type Ia Supernovae Observations}

\author{L. R. Cola\c{c}o}
\email{colacolrc@gmail.com}
\affiliation{Instituto Federal de Rondônia, 76850-000, Guajará-Mirim - RO, Brazil}
\affiliation{Departamento de F\'{\i}sica, Universidade Federal de Campina Grande, 58429-900, Campina Grande - PB, Brazil}

\author{R. F. L. Holanda}
\email{holandarfl@gmail.com}
\affiliation{Universidade Federal do Rio Grande do Norte, Departamento de F\'{i}sica Te\'{o}rica e Experimental, 59300-000, Natal - RN, Brazil.}
\affiliation{Departamento de F\'{\i}sica, Universidade Federal de Campina Grande, 58429-900, Campina Grande - PB, Brazil}

\author{Marcelo Ferreira}
\email{fsm.fisica@gmail.com}
\affiliation{Universidade Federal do Rio Grande do Norte, Departamento de F\'{i}sica Te\'{o}rica e Experimental, 59300-000, Natal - RN, Brazil.}

\begin{abstract}
\noindent 
{  In this paper, we investigate a possible time variation of the gravitational constant ($G$) using a non-parametric approach. Our main cosmological probe is the gas mass fraction of galaxy clusters measured from X-ray observations. We also account for the effect of a varying $G$ on the intrinsic luminosity of type Ia supernovae (SNe Ia) through the Chandrasekhar mass–luminosity relation. We consider a specific phenomenological scenario, motivated by some scalar–tensor and screened modified-gravity frameworks, in which the standardized luminosity of SNe Ia decreases with increasing Chandrasekhar mass. Using gas mass fraction measurements jointly with luminosity distances from the Pantheon+ compilation, we reconstruct the evolution of $G$ through Gaussian Processes. Our results indicate that a constant gravitational coupling remains broadly consistent with the data, although mild low-redshift departures are allowed.
}

\end{abstract}



\maketitle

\section{Introduction}

An intriguing possibility, first proposed by Paul Dirac in 1937 \cite{Dirac1937ti}, is that the fundamental constants of nature—such as the fine-structure constant ($\alpha$) or the gravitational constant ($G$)—might not be truly constant but could vary over cosmological timescales. Detecting or constraining such variations would have profound implications for our understanding of fundamental interactions, potentially pointing toward new physics beyond General Relativity and Quantum Field Theory \cite{Uzan2002vq,Uzan2010pm,Martins2017yxk,Uzan2024ded}. 
Despite its fundamental role in physics, $G$ remains the least precisely determined among the fundamental constants. Laboratory experiments—using different techniques such as torsion balances \cite{Gundlach2000rk,101093nsrnwaa165}, pendulums \cite{PhysRevLett852869}, and more recently cold-atom interferometry \cite{Tino2020dsl}—still yield values that differ beyond their quoted uncertainties, indicating the presence of unresolved systematic effects. Thanks to recent advances in astrophysical observations, cosmological probes now provide complementary ways to test the possible evolution of $G$ over cosmic time. These include measurements from the cosmic microwave background (CMB) \cite{Umilta2015cta,Ballardini2016cvy,Bai2015vca,Xue2014kna}, big-bang nucleosynthesis (BBN) \cite{Alvey2019ctk,Gelmini2020ekg}, Type Ia supernovae (SNe Ia) \cite{Zhang2017aqn,Wright2017rsu,Zhao2018gwk}, gravitational-wave observations \cite{Zhao2018gwk,Vijaykumar2020nzc}, and strong gravitational lensing \cite{Holanda2025xsj}. These probes allow us to constrain the strength of gravity at different epochs of the Universe’s evolution (see \cite{2003RvMP...75..403U,Uzan2011} for excellent reviews).
 
{  Particularly for Type Ia supernovae (SNe Ia), the intrinsic luminosity depends on the Chandrasekhar mass of the white dwarf progenitor, which scales as \(M_{\rm Ch}\propto G^{-3/2}\) \cite{Amendola1999vu,GarciaBerro1999cwy,gaztanaga2001bounds,Mould2014iga,Ruchika2023ugh}. Therefore, variations in the gravitational coupling may modify the peak luminosity and light-curve properties of SNe Ia, potentially introducing a redshift dependence that challenges the assumption that they are standardizable candles. Such effects may arise in some modified-gravity scenarios, including scalar--tensor theories, screened gravity models, and viable \(f(R)\) frameworks, where the effective gravitational coupling evolves with cosmic time, \(G_{\textrm{eff}}=G_{\textrm{eff}}(z)\) \cite{PhysRevD66023525,LorenAguilar2003qtx,Brax2008hh,CeronHurtado2016jrp,Khoury2003rn,Wang2012kj,Khoury2003aq}. Within this context, semi-analytic studies of SNe Ia light curves \cite{Wright2017rsu,Sakstein2019qgn} have suggested a phenomenological scaling relation of the form \(L_{\textrm{SNe}} \propto M_{\textrm{Ch}}^{-0.97}\), implying \(L_{\textrm{SNe}} \propto G^{1.46}\). Such phenomenological behavior may emerge from the combined effects of modified gravity, white dwarf structure, and explosion dynamics.}

In recent years, several studies have explored the cosmological consequences of a varying $G$. For instance, Ref.~\cite{Holanda2025xsj} tested parameterizations of the form $G(z)=G_0(1+G_1 z)$ and $G(z)=G_0(1+z)^{G_1}$, taking into account their impact on the luminosity of SNe Ia through the Chandrasekhar mass–luminosity relation, jointly with gravitational lensing observations. In \cite{Ruchika2023ugh}, the authors proposed a gravitational transition scenario in which $G$ is modified on scales below $\approx 50$ Mpc, leading to changes in the Cepheid Period–Luminosity relation and in the intrinsic brightness of SNe Ia. With a fractional variation of $\Delta G/G \approx 0.04$, the model was able to reproduce a value of the Hubble constant ($H_0$) consistent with the Planck measurement. More recently, Ref.~\cite{Lamine2024xno} employed the latest Planck PR4 data combined with DESI BAO observations to constrain time-varying $G$, finding no statistically significant evidence for this hypothesis. Their conclusions remained stable under a wide range of cosmological assumptions, including non-flat geometries and exotic dark-energy components. However, the situation may change when variations in both the fine-structure constant ($\alpha$) and $G$ are considered simultaneously. In Ref.~\cite{BezerraSobrinho2025vaf}, observational data from SNe Ia, BAO, and CMB were used to constrain possible co-variations of $G$ and $c$. By adopting different parametrizations for $c(z)$ motivated by varying speed of light (VSL) scenarios, the Pantheon$+$ sample suggested a variable speed of light at more than $3\sigma$ confidence level. This apparent discrepancy arises from a strong correlation between $H_0$ and the VSL parameter.

{  In this work, we show that combining measurements of $(f_{\textrm{gas}}^{\textrm{X-ray}})$ with luminosity distance data from SNe Ia $((D_{L}^{\textrm{SNe}}))$ provides  a complementary test for a possible time variation of the gravitational constant. Both SNe Ia and the gas mass fraction of galaxy clusters depend on the local value of the Newtonian gravitational constant, making them suitable probes to test a possible cosmic evolution of gravity. In the case of SNe Ia, variations in $G$ modify the Chandrasekhar mass of the white dwarf progenitors, affecting their intrinsic luminosities and consequently the inferred luminosity distances. Likewise, measurements of the gas mass fraction of galaxy clusters obtained from X-ray observations are sensitive to $G$ through its influence on cluster dynamics and total mass estimates. In this analysis, we focus on a specific phenomenological scenario, motivated by some scalar–tensor and screened modified-gravity frameworks, in which the standardized luminosity of SNe Ia decreases with increasing Chandrasekhar mass. We allow for a possible redshift evolution of the gravitational coupling, parameterized as ($G(z)=G_0\phi(z)$), and reconstruct the function $(\phi(z))$ using a non-parametric approach. For the X-ray data, we consider the updated sample of $(f_{\textrm{gas}}^{\textrm{X-ray}})$ compiled by \cite{101093mnrasstab3390}, which includes 44 massive, hot, and morphologically relaxed galaxy clusters in the redshift range $(0.018 \leq z \leq 1.160)$ observed by Chandra. As a complementary probe, we use luminosity distances from SNe Ia drawn from the Pantheon(+) compilation \cite{Scolnic2021amr}.}

This paper is organized as follows. The methodology adopted in this work is presented in section \ref{Methodology}. In section \ref{samples}, we describe the cosmological data used in our analysis. The analysis and main results are presented in section \ref{results}. Finally, we conclude in section \ref{final}.

\section{Methodology}
\label{Methodology}

In order to connect a possible variation of the gravitational constant $G$ with observable quantities, we first describe how the gas mass fraction of galaxy clusters depends on $G$, and then discuss how changes in $G$ affect the intrinsic luminosity of Type Ia supernovae (SNe Ia). Finally, we discuss how these two observables can be combined to place constraints on possible variations of $G$.

\subsection{Gas Mass Fraction} \label{sec.gas mass}

The baryonic matter content of galaxy clusters is predominantly composed of intracluster gas, a diffuse hot gas that emits mainly in the X-ray band through the thermal bremsstrahlung process \cite{2011ARAA49409A,2017PhRvD95h4006H,1996PASJ48L119S,101093mnrasstab3390}. In this context, a cosmologically significant parameter is the gas mass fraction, defined as the ratio between the intracluster gas mass and the total mass of the galaxy cluster:

\begin{equation} \label{fg}
    f_{\mathrm{gas}} \equiv \frac{M_{\mathrm{gas}}}{M_{\mathrm{tot}}},
\end{equation}
where the total mass consists of baryonic and dark matter. 

Assuming a spherically symmetric intracluster gas in hydrostatic equilibrium, described by an isothermal $\beta$-model, the gas mass inferred from X-ray observations is given by (see more details in \cite{sarazin}):

\begin{eqnarray} \label{mgas}
M_{\mathrm{gas}}(<R) &=&
\left( \frac{3 \pi \hbar m_e c^2}{2(1+X)e^2} \right)^{1/2}
\left( \frac{3 m_e c^2}{2 \pi k_B T_e} \right)^{1/4}
m_H
\frac{r_c^{3/2}}{g_B (T_e)^{1/2}}
\nonumber \\
&\times&
\left[
\frac{I_M(R/r_c, \beta)}{I_L^{1/2}(R/r_c, \beta)}
\right]
[L_X(<R)]^{1/2},
\end{eqnarray}
where $X$ denotes the hydrogen mass fraction, $m_e$ is the electron mass, $r_c$ is the core radius, $g_B(T_e)$ is the Gaunt factor, $c$ is the speed of light, $\hbar$ is the reduced Planck's constant, $e$ is the electrical charge, $k_B$ is the Boltzmann constant, $m_H$ is the hydrogen mass, and $L_X(<R)$ is the total X-ray luminosity. The dimensionless integrals $I_M$ and $I_L$ are defined, respectively, as

\begin{equation}
I_M(y, \beta) \equiv \int_0^y (1 + x^2)^{-3\beta/2} x^2 \, dx,
\end{equation}

\begin{equation}
I_L(y, \beta) \equiv \int_0^y (1 + x^2)^{-3\beta} x^2 \, dx.
\end{equation}

\noindent On the other hand, the total mass enclosed within a radius $R$ can be determined under the assumption of hydrostatic equilibrium \cite{2011ARAA49409A}:

\begin{equation}\label{mtot}
M_{\mathrm{tot}}(<R) = -\frac{k_B T_e R}{G \mu m_H}
\left. \frac{d \ln n_e(r)}{d \ln r} \right|_{r=R},
\end{equation}
where $T_e$ and $n_e$ denote the electron temperature and density, respectively, while $\mu$ is the mean molecular weight.

Therefore, combining Eq.s (\ref{mgas}) and (\ref{mtot}), it follows that $f_{\mathrm{gas}} \propto G$. Consequently, if we allow the gravitational constant to vary with redshift as $G = G_0 \phi(z)$, where $\phi(z)$ represents a dimensionless function describing the possible redshift dependence of $G$, Eq. (\ref{fg}) must be corrected accordingly as:

\begin{equation}\label{fgas0}
    f_{\mathrm{gas}}(z)=\phi(z)f_{\mathrm{gas},0} \,,
\end{equation}
where $f_{\mathrm{gas},0}$ is the gas mass fraction obtained with the current value of $G$.

As a cosmological probe, studies using the gas mass fraction are achieved using the following equation \cite{allen2008improved,Ettori2009,2011ARAA49409A,Mantz2014xba,Holanda_2019,101093mnrasstab3390}:

\begin{equation}
\label{fgas1}
f_{\mathrm{gas}}(z) = \gamma_g(z)K(z) A(z) \left[\frac{\Omega_b}{\Omega_m}\right] \left(\frac{D_L^*}{D_L}\right)^{3/2},
\end{equation}
where $\gamma_g(z)$ corresponds to the gas depletion parameter that quantifies the fraction of baryonic gas that is thermalized within the cluster potential \cite{allen2008improved,Ettori2009,2011ARAA49409A,Mantz2014xba}. The term $K(z)$ represents the mass calibration factor, while $A(z)$ is the angular correction factor, which remains close to unity for all relevant cosmologies and redshifts and can therefore be neglected without introducing significant systematic error \cite{allen2008improved}. The quantity $D_L^*$ denotes the luminosity distance to the galaxy cluster adopted in the observations to determine $f_{\mathrm{gas}}$ (typically a flat $\Lambda$CDM model with $H_0 = 70~\mathrm{km~s^{-1}~Mpc^{-1}}$ and $\Omega_m = 0.3$ is used). It is worth noting that cosmological analyses based on gas mass fraction measurements are effectively model-independent because of the ratio in parentheses in Eq.(\ref{fgas1}), which accounts for the expected variation in the measured $f_{\mathrm{gas}}$ when the underlying cosmological model is changed.

Although the gas mass fraction in galaxy clusters is a powerful cosmological probe, its use relies on assumptions such as hydrostatic equilibrium and the stability of the baryon fraction, which may introduce systematic uncertainties. Nevertheless, the relevant astrophysical parameters describing cluster physics have been extensively investigated in the literature, with well-established ranges of values \cite{Mantz2014xba,applegate2014weighing,101093mnrasstab3390}. Therefore, these aspects must be carefully considered when using this observable to constrain possible variations of $G$. Following \cite{101093mnrasstab3390}, we allow for a possible redshift and mass dependence of $\gamma_g(z)$ through parameterization  

\begin{equation}
    \gamma_g(z) = \gamma_0 \,(1 + \gamma_1 z) 
    \left( \frac{M_{2500}}{3 \times 10^{14} \, M_{\odot}} \right)^{\alpha},
\end{equation}
where $\alpha = 0.025 \pm 0.033$ \cite{101093mnrasstab3390} and $M_{2500}$ is the total mass within the radii of interest ($r_{2500}$). Similarly, we adopt the redshift-dependent of $K(z)$ as $K(z) = K_0 (1 + K_1 z)$. Consistently with \cite{Mantz:2014xba,101093mnrasstab3390}, we apply the Gaussian priors: $\gamma_0 = 0.79 \pm 0.07$ and $-0.05 \geq \gamma_1 \leq 0.05$ \cite{2013ApJ777123B,Battaglia2013,planelles2013baryon,Holanda2017}, and $K_0 = 0.93 \pm 0.11$ and $-0.05 \geq K_1  \leq 0.05$ \cite{applegate2014weighing}.  
    {  The depletion factor parameter is derived from hydrodynamical simulations and corresponds directly to the gas mass fraction (see Table III in \cite{planelles2013baryon} and also \cite{Holanda2017,Mantz:2014xba,101093mnrasstab3390}), while $K_0$ is obtained from an analysis of 13 clusters that includes weak-lensing mass measurements from the \textit{Weighing the Giants} project \cite{applegate2014weighing}. The ratio $\Omega_b/\Omega_m$ entering Eq.~(\ref{fgas1}) is adopted from the most recent galaxy-clustering determination reported in Ref.~\cite{2025PhRvD.111f3526K}. We use this external prior in order to avoid imposing a specific CMB-derived background cosmological model. The uncertainties associated with these astrophysical and cosmological nuisance quantities are propagated in the Monte Carlo analysis.}

However, to employ the luminosity distances from Type Ia supernovae in Eq. (\ref{fgas1}) to constrain $\phi(z)$, it is necessary to examine how possible evolution of the gravitational constant would affect the intrinsic luminosity of SNe Ia. The following discussion explores this dependence in detail.

\subsection{Chandrasekhar mass and
the luminosity of Type Ia supernovae}

To determine the luminosity distances to galaxy clusters using SNe Ia, it is first necessary to understand how a possible evolution of $G$ influences their intrinsic luminosities. It is well-known that SNe Ia serve as standard candles in cosmology owing to the well-established correlation between their intrinsic luminosities and light-curve properties. However, any evolution in fundamental physical constants, such as $G$ or $\alpha$, may impact cosmological distance measurements. In this context, we investigate how variations in $G$ affect the intrinsic luminosity of SNe Ia, particularly through its dependence on the Chandrasekhar mass. Although it was traditionally assumed that the luminosity of SNe Ia increases with Chandrasekhar mass $M_{\mathrm{Ch}}$, the exact nature of this relationship remains under debate. Recent studies have suggested that intrinsic luminosity may exhibit an inverse dependence on $M_{\mathrm{Ch}}$ in non-standard gravity theories.

According to several authors, the peak luminosity of SNe Ia is proportional to the Chandrasekhar mass of exploding white dwarfs by $L_{\mathrm{SNe}} \propto M_{\mathrm{Ch}}$ \cite{woosley1986physics,Amendola1999vu,gaztanaga2001bounds}. Since the Chandrasekhar mass scales as $M_{\mathrm{Ch}} \propto G^{-3/2}$, such a relation implies an inverse dependence between the gravitational constant $G$ and both $M_{\mathrm{Ch}}$ and $L_{\mathrm{SNe}}$. Consequently, an increase in $G$ would lead to a decrease in their luminosity. Thus, the measured luminosity distance from SNe Ia should be corrected by $D_L^{\mathrm{SNe}} = D_{L,0}^{\mathrm{SNe}} \phi(z)^{-3/4}$, where $D_{L,0}^{\mathrm{SNe}}$ represents the luminosity distance measured under the assumption of a constant $G$.

\begin{figure*}[htbp]
\includegraphics[width=0.49\textwidth]{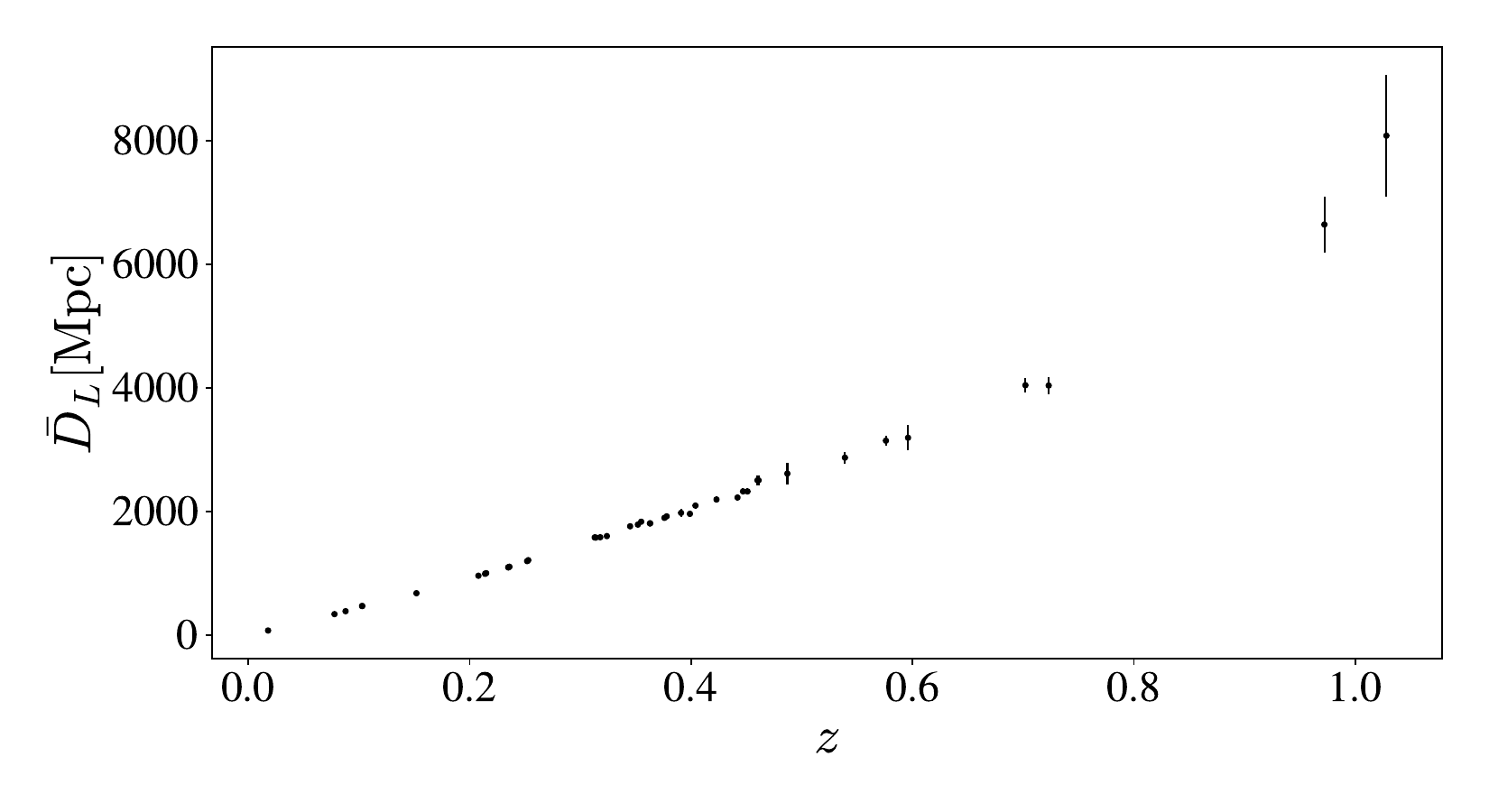}
\includegraphics[width=0.49\textwidth]{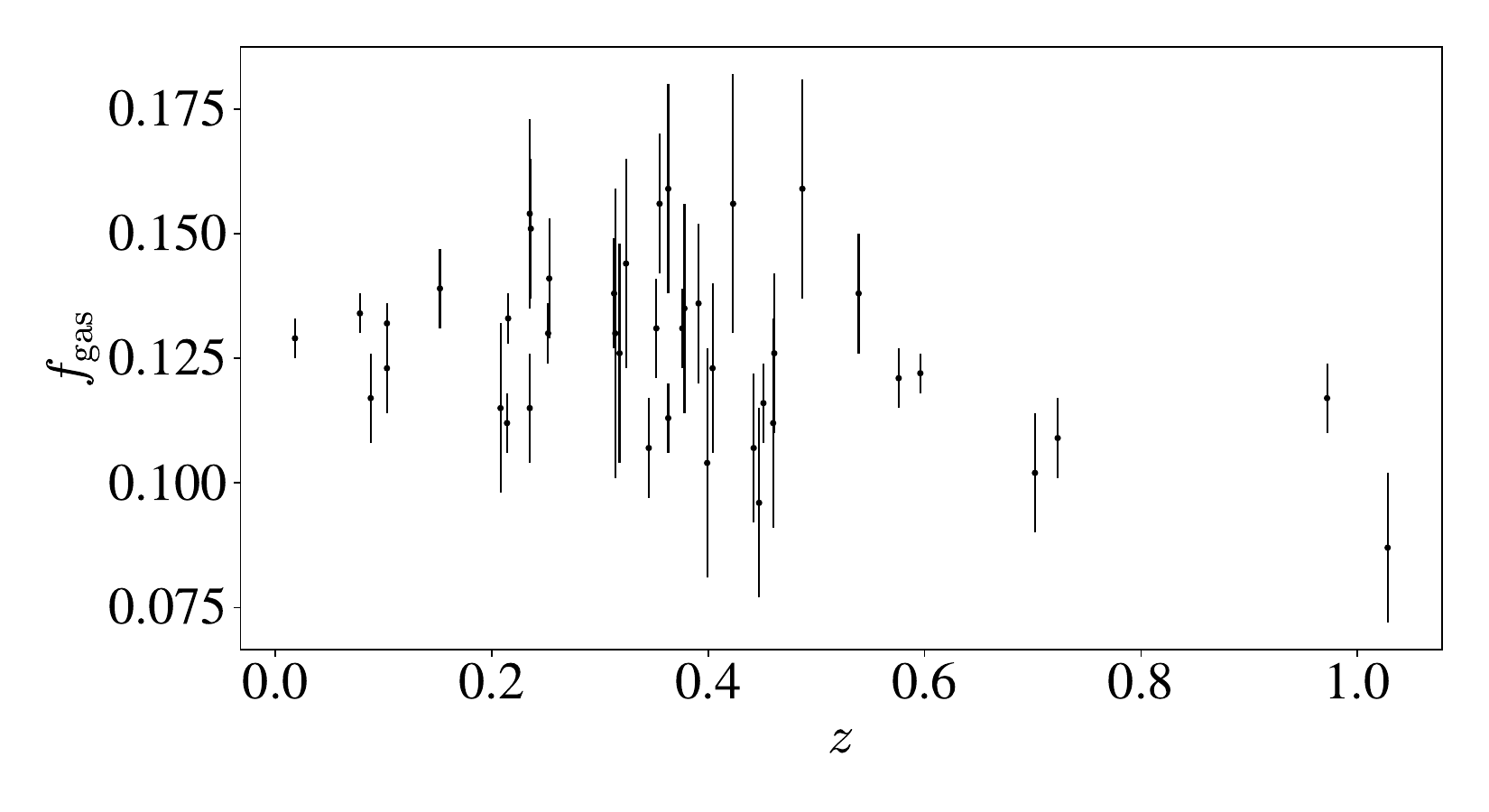}
\caption{Left Panel: The weighted-average luminosity distances from Pantheon$+$ dataset \cite{Scolnic2021amr} for each galaxy cluster. Right Panel: The selected gas mass fraction measurements from \cite{101093mnrasstab3390}.}
\label{snefgas}
\end{figure*}

{  On the other hand, in some non-standard gravity frameworks, the gravitational coupling is associated with a scalar field whose value may evolve in space and time \cite{Amendola1999vu,GarciaBerro1999cwy,PhysRevD66023525,Brax2008hh,Wang2012kj,CeronHurtado2016jrp,Khoury2003aq,Khoury2003rn}. In such scenarios, the effective gravitational coupling $G_{\textrm{eff}}(z)$ may vary with redshift, potentially modifying both the cosmic expansion history and the structure of white dwarf progenitors \cite{Brans:1961sx,Li2015aug,Hamity2005ri,GarciaBerro1999cwy}. Since the Chandrasekhar mass scales as $M_{\textrm{Ch}}\propto G^{-3/2}$, variations in $G$ may affect the explosion properties and luminosities of SNe Ia.
Semi-analytic studies of SNe Ia light curves \cite{Wright2017rsu,Sakstein2019qgn} have suggested that the standardized peak luminosity does not necessarily increase with Chandrasekhar mass. In these scenarios, the luminosity depends not only on the total ejecta mass, but also on the balance between thermonuclear energy production and radiative diffusion within the expanding material \cite{Wright2017rsu,Gupta2021tma,Ballardini2021eox}. As a result, weaker gravity may lead to larger Chandrasekhar masses while simultaneously producing less luminous SNe Ia. Such a behavior can emerge from the combined effects of modified gravity, white dwarf structure, and explosion dynamics. In particular, the semi-analytic model developed by \cite{Wright2017rsu} (see also \cite{2019PhRvD.100j4035S}) suggests a phenomenological scaling relation of the form $L_{\mathrm{SNe}} \propto M_{\mathrm{Ch}}^{-0.97}$. Combining this with $M_{\mathrm{Ch}} \propto G^{-3/2}$, one obtains $L_{\mathrm{SNe}} \propto G^{1.46}$. Assuming that the evolution of $G$ directly influences both $M_{\mathrm{Ch}}$ and $L_{\mathrm{SNe}}$, we may write}

\begin{equation}
D_L^{\mathrm{SNe}} = D_{L,0}^{\mathrm{SNe}} \phi(z)^{0.73}.
\end{equation}
Taking into account these modifications in the luminosity of SNe Ia, Eq. (\ref{fgas1}) becomes

\begin{equation} \label{fgas_correct}
f_{\mathrm{gas}}(z) = \gamma_g(z)K(z)
\left[\frac{\Omega_b}{\Omega_m}\right]
\left(\frac{D_L^*}{D_{L,0}\phi(z)^n}\right)^{3/2},
\end{equation}
where $n = -3/4$ corresponds to the standard gravity scenario, and $n = 0.73$ to the non-standard gravity one. 
Thus, combining this expression with Eq. (\ref{fgas0}), we may obtain:

\begin{equation} \label{phi}
    \phi^{(3n + 2)/2} (z)=
    \frac{K(z)\,\gamma_g(z)\,(\Omega_b / \Omega_m)}{f_{\mathrm{gas},0}}
    \left(\frac{D_L^*}{D_{L,0}}\right)^{3/2}.
\end{equation}
{ This key equation, Eq.~(\ref{phi}), is used to reconstruct $\phi(z)$ from observational data. We note that the standard scenario corresponding to $(n=-3/4)$ leads to a highly nonlinear dependence of $\phi(z)$ on the observational quantities, which amplifies statistical fluctuations and results in unstable and poorly constrained reconstructions. For this reason, in the present analysis we restrict our study to the non-standard phenomenological scenario discussed above.}

\section{Samples} \label{samples}

\begin{itemize}

\item{ \bf  Gas mass fraction:}
We utilize a recent sample of $f_{\mathrm{gas}}$ measurements spanning the redshift range $0.018 \leq z \leq 1.160$ and compiled by \cite{101093mnrasstab3390}. This dataset consists of 44 massive, hot, and morphologically relaxed galaxy clusters observed by the \textit{Chandra} X-ray Observatory (see Fig. \ref{snefgas}). The selection of relaxed systems is crucial for minimizing systematic uncertainties and intrinsic scatter associated with deviations from hydrostatic equilibrium and spherical symmetry. Furthermore, the gas mass fractions were determined within spherical shells at radii near $r_{2500}$ ($0.8$--$1.2\,r_{2500}$), where $r_{2500}$ denotes the radius within which the mean enclosed mass density equals $2500$ times the critical density of the Universe at the cluster redshift.

\item{\bf SNe Ia Sample:}

We employ luminosity distances from the largest compiled dataset of SNe Ia observations, called Pantheon+ \cite{Scolnic2021amr}. This compilation contains 1701 light curves of 1550 spectroscopically confirmed SNe Ia spanning the redshift range $0.001 \leq z \leq 2.26$. Thus, the luminosity distance for each supernova is obtained from the relation:

\begin{equation}\label{SNeIa}
    D_L = 10^{(m_b - M_B - 25)/5} \ \text{Mpc},
\end{equation}
where $m_b$ and $M_B$ are the apparent and the absolute magnitudes, respectively. In this paper, we consider the absolute local estimate from the SH0ES team \cite{Scolnic2021amr,riess2022comprehensive} in order to perform an independent cosmological model test.

The Pantheon$+$ dataset also provides the covariance matrix of $m_b$ for each supernova, which must be propagated into the covariance matrix of $D_L$ through the relation \cite{Ferreira2024kht}:

\begin{eqnarray}
\text{cov}({\bf D_L, D_L}) = \left(  \frac{\partial {\bf D_L}}{\partial {\bf m_b}} \right) \text{cov}({\bf m_b, m_b}) \nonumber \\
         \times \left(  \frac{\partial {\bf D_L}}{\partial {\bf m_b}} \right)^T,
     \end{eqnarray}
where $\left(  \frac{\partial {\bf D_L}}{\partial {\bf m_b}} \right)$ is the Jacobian matrix of the transformation, and the variables in bold correspond to the vector representations of each data set.

To perform our analyses, we must match SNe Ia and galaxy clusters within identical redshift bins. For each galaxy cluster, we select SNe Ia whose redshifts satisfy $\lvert z_{\mathrm{GC}} - z_{\mathrm{SN}} \rvert \leq 0.005$. The corresponding SNe Ia luminosity distances are then combined through the following weighted average, which is constructed using the full covariance matrix in order to account for uncertainties and correlations among the data properly:

\begin{equation}
\label{DL_mean}
\bar{D}_L = 
\frac{\sum_{i,j} D_{L_i}\, w_{ij}}
     {\sum_{i,j} w_{ij}},
\end{equation}
where $D_{L_i}$ denotes the luminosity distance of the $i$-th selected supernova, and  $w_{ij} = \left[\mathrm{cov}^{-1}(\mathbf{D}_L, \mathbf{D}_L)\right]_{ij}$ are the elements of the inverse covariance matrix. This procedure yields a total of 42 measurements of $\bar{D}_L$, where each corresponds to a cluster of our sample \cite{101093mnrasstab3390} (see Fig. \ref{snefgas}). Furthermore, we note that the weighted-average luminosity distances are not independent, as their construction inherits the intrinsic correlations present in the supernova data, as well as the correlations induced by the overlap of SNe Ia contributing to distinct $\bar{D}_L$ estimates. Thus, we may demonstrate that the covariance matrix of the averaged luminosity distances satisfies the relation \cite{colacco2022varying}:

\begin{equation*}\label{covDL}
\text{cov}(\bar{D}_{L}^i, \bar{D}_{L}^j) =
\end{equation*}
\begin{equation}
\sum_{\alpha,\gamma}^{n^i, n^j} \frac{\text{cov}(D^i_{L_\alpha}, D^j_{L_\gamma}) \left[ \sum_{\beta}^{n^i} w_{\alpha \beta}^i \right] \left[ \sum_{\sigma}^{n^j} w_{\gamma \sigma}^j \right]}{\left[ \sum_{\sigma} ^{n^i} \sum_{\beta}^{n^i} w_{\sigma \beta}^i \right] \left[ \sum_{\sigma} ^{n^j} \sum_{\beta}^{n^j} w_{\sigma \beta} ^j\right]},
\end{equation}
where the superior indices represent the sets of SNe considered to calculate the luminosity distances in Eq. (\ref{DL_mean}), whereas the lower indices correspond to the $i$-th term in each set, and $n$ is the number of data points.
\end{itemize}

\section{Analysis and Main Results}
\label{results}

In this work, we aim to reconstruct the function $\phi(z)$ using Gaussian Processes (GPs), a non-parametric statistical approach widely applied in regression and classification analyses. To apply the GP to Eq. (\ref{phi}), we first transform the previously described observational data sets into mean estimates of $\phi(z)$ along with their corresponding covariance matrix. Each observable is modeled as a Gaussian random variable, and its uncertainty is propagated to $\phi(z)$ through Monte Carlo sampling. This procedure is necessary because the gas mass fraction data exhibit large uncertainties, rendering standard Gaussian error propagation based on partial derivatives inadequate. { The resulting data points are presented in Table~\ref{tab:phi_values}, where the uncertainties were obtained from the diagonal elements of the covariance matrix.}

\begin{figure*}[htb]
    \centering
        \centering
        \includegraphics[width=0.75\textwidth]{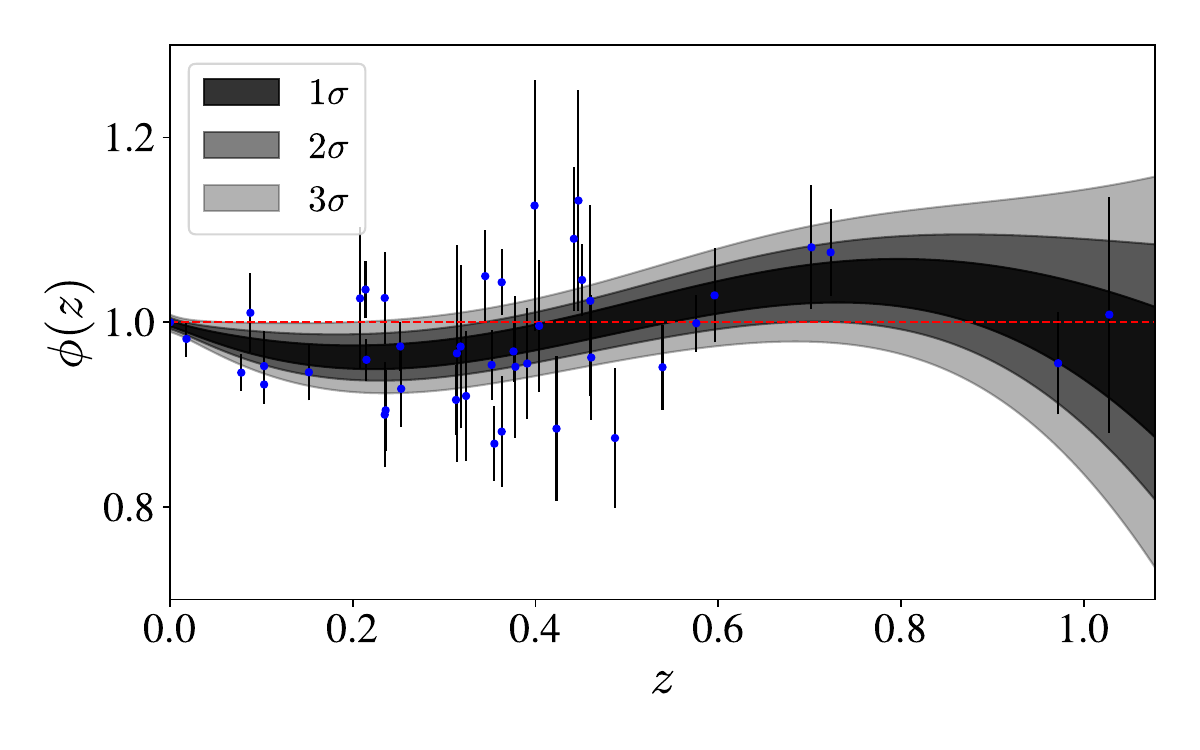}
    \vspace{0.5cm} 
    \caption{
    reconstruction of $\phi(z)$ in the non-standard framework ($n = 0.73$).  
    The dashed line represents the scenario with no variation in $G$. }
    \label{reconstructions}
\end{figure*}

For the SNe Ia data, whose measurements exhibit non-negligible correlations, we model them using a multivariate Gaussian distribution that incorporates the full covariance matrix. For the other observables, we assume independent one-dimensional Gaussian distributions. GPs offer a non-parametric approach to reconstructing a smooth function directly from the data, without assuming any specific functional form. At each redshift $z$, the reconstructed quantity $\phi(z)$ follows a Gaussian distribution, and correlations between values at different redshifts are encoded through a covariance function $k(z,z')$. Thus, implementing a GP requires specifying a prior mean function and choice of kernel (see \cite{Seikel:2012uu} for further details). In our analyses, we adopt a zero prior mean function to avoid introducing biases, and we use the standard Gaussian kernel as the covariance function between two data points separated by a redshift interval $z-z'$:

\begin{equation} \label{kernel}
    k(z,z') = \sigma_f^2 \exp \left( -  \frac{(z - z')^2}{2l^2} \right).
\end{equation}
The hyperparameter $l$ characterizes the characteristic redshift scale in which the function exhibits significant variations, while $\sigma_f$ sets the typical amplitude of these variations. We also impose the boundary condition $G(z=0)=G_0$ { (equivalently, $\phi(z = 0) = 1$). To impose this condition, we include an additional data point at $z = 0$ consistent with $\phi(0) = 1$ (Table \ref{tab:phi_values}). Nevertheless, in order to avoid introducing potential biases in the reconstruction, we adopt a zero prior mean function \cite{Seikel}}. 
The hyperparameters are determined by maximizing the GP log-likelihood:

\begin{eqnarray} \label{likelihood_GP} \ln {\mathcal{L}}=-\frac{1}{2} \bm {\bm (\phi)}^T [\bm K(\bm {z}, \bm {z}) + \bm C]^{-1} \bm {\bm (\phi)} \nonumber \\ - \frac{1}{2} \ln | \bm K(\bm z, \bm z) + \bm C| - \frac{\mathrm{n}}{2} \ln 2 \pi, \end{eqnarray}
where $\bm{\phi}$ and $\bm{z}$ denote the vectors of dependent and independent variables, respectively. The matrix $\bm{K}(\bm{z},\bm{z})$ corresponds to the GP covariance matrix whose elements are computed using Eq. (\ref{kernel}), while $\bm{C}$ denotes the observational covariance matrix that contains the uncertainties of the reconstructed function. The quantity $\mathrm{n}$ refers to the total number of data points used in our analyses. The GP reconstruction is performed with the GaPP package\footnote{\url{https://github.com/carlosandrepaes/GaPP}}.

\begin{table*}[htb!]
\resizebox{\textwidth}{!}{%
\setcounter{table}{0} 
\begin{tabular}{l c c c}
\hline
\textbf{Model} & $G(z = 0.3)\times 10^{-11}$ & $G(z = 0.6)\times 10^{-11}$ & $G(z = 0.9)\times 10^{-11}$ \\ \hline
Non-standard SNe Ia Scenario ($n = 0.73)$ & $(6.4573 \pm 0.0904)$ & $(6.8552 \pm 0.1174)$ & $(6.8507 \pm 0.2328)$ \\ 
\hline
\end{tabular} 
}
\caption{Predicted values for $G(z)$ at three different $z$ values at a $1 \sigma$ confidence level. $G$ is calculated in $\mathrm{m^3\,kg^{-1}\,s^{-2}}$ units.}
\label{tabela}
\end{table*}

{ We utilized the locally determined prior on $M_B$ from the SH0ES team, $M_B = - 19.253 \pm 0.027$ \cite{riess2022comprehensive}, to obtain $D_L(z)$ and its associated covariance matrix, along with the gas mass fraction sample calculated at $r_{2500}$ from \cite{101093mnrasstab3390}, to derive $\phi(z)$ and its corresponding covariance matrix using the Eq. \eqref{phi} through a Monte Carlo sampling. The priors on the astrophysical quantities of the galaxy clusters, namely $\gamma(z), K(z), A(z)$, and also $\Omega_b / \Omega_m$ are described in Sec. \ref{sec.gas mass}, and their errors are propagated in the analysis via Monte Carlo sampling.}  
The results are presented in Fig. \ref{reconstructions}. Our results indicate that, within the specific phenomenological scenario considered in this work ($n=0.73$), the reconstructed evolution of $G(z)$ is compatible with a constant gravitational coupling over the redshift range analyzed. Although the reconstruction allows mild low-redshift departures from a constant $G$ ($z<0.4$), no statistically significant deviation is observed. Nevertheless, such scenarios may still have interesting implications for cosmological analyses involving varying gravitational couplings.

\begin{table*}[ht]
\centering
\caption{Values of $\phi(z)$ and their corresponding $1\sigma$ uncertainties.}
\begin{tabular}{ccc|ccc|ccc}
\hline
$z$ & $\phi(z)$ & $\sigma_{\phi}$ &
$z$ & $\phi(z)$ & $\sigma_{\phi}$ &
$z$ & $\phi(z)$ & $\sigma_{\phi}$ \\
\hline

0.000 & 1.0000 & 0.0003 &
0.253 & 0.9281 & 0.0410 &
0.423 & 0.8849 & 0.0782 \\

0.018 & 0.9820 & 0.0192 &
0.313 & 0.9160 & 0.0378 &
0.442 & 1.0904 & 0.0778 \\

0.078 & 0.9455 & 0.0197 &
0.314 & 0.9664 & 0.1175 &
0.447 & 1.1317 & 0.1199 \\

0.088 & 1.0103 & 0.0428 &
0.318 & 0.9737 & 0.0885 &
0.451 & 1.0458 & 0.0392 \\

0.103 & 0.9327 & 0.0207 &
0.324 & 0.9203 & 0.0700 &
0.460 & 1.0233 & 0.1032 \\

0.103 & 0.9526 & 0.0378 &
0.345 & 1.0499 & 0.0501 &
0.461 & 0.9619 & 0.0673 \\

0.152 & 0.9460 & 0.0303 &
0.352 & 0.9539 & 0.0378 &
0.487 & 0.8748 & 0.0753 \\

0.208 & 1.0259 & 0.0767 &
0.355 & 0.8687 & 0.0404 &
0.539 & 0.9514 & 0.0462 \\

0.214 & 1.0354 & 0.0313 &
0.363 & 0.8817 & 0.0597 &
0.576 & 0.9989 & 0.0307 \\

0.215 & 0.9594 & 0.0228 &
0.363 & 1.0433 & 0.0358 &
0.596 & 1.0291 & 0.0508 \\

0.235 & 0.9001 & 0.0567 &
0.376 & 0.9685 & 0.0329 &
0.702 & 1.0811 & 0.0669 \\

0.235 & 1.0263 & 0.0497 &
0.378 & 0.9518 & 0.0765 &
0.723 & 1.0757 & 0.0471 \\

0.236 & 0.9048 & 0.0437 &
0.391 & 0.9554 & 0.0601 &
0.972 & 0.9557 & 0.0554 \\

0.252 & 0.9739 & 0.0265 &
0.399 & 1.1263 & 0.1357 &
1.028 & 1.0083 & 0.1277 \\

0.404 & 0.9961 & 0.0716 &
& & \\

\hline
\end{tabular}
\label{tab:phi_values}
\end{table*}

It is important to stress that Ref.~\cite{Ruchika2024ymt} analyzed the SH0ES’22 sample, including 37 Cepheid calibrator galaxies and 42 Type Ia supernovae. By modeling a late-time transition in the gravitational constant $G$, which affects both Cepheid distances and supernova luminosities, the authors fitted the relation $L \propto M_c^n$ and obtained $n = -1.68 \pm 0.68$, indicating a non-standard inverse relation between luminosity and Chandrasekhar mass. They also reported mild evidence for a transition in $G$ at a distance of $\sim 22.4\,\mathrm{Mpc}$ (about $73$ Myr ago), with a value of the Hubble constant consistent with CMB measurements if $G$ was $\sim 4\%$ larger in the past. Very recently, Ref.~\cite{2026PhRvD.113b3501P} revisited this scenario and reinforced these results. Together with our findings, these studies highlight the importance of exploring possible variations of $G$ in cosmological parameter inference.

\begin{figure*}[htbp]
\includegraphics[width=0.49\textwidth]{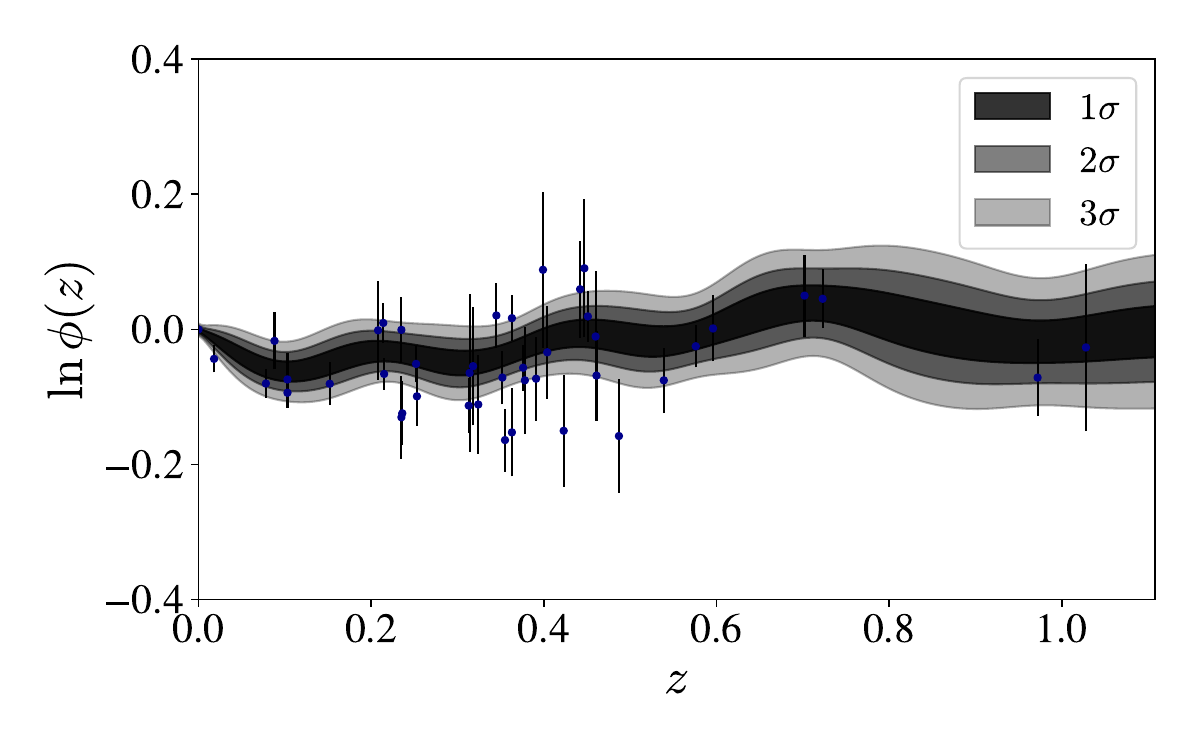}
\includegraphics[width=0.49\textwidth]{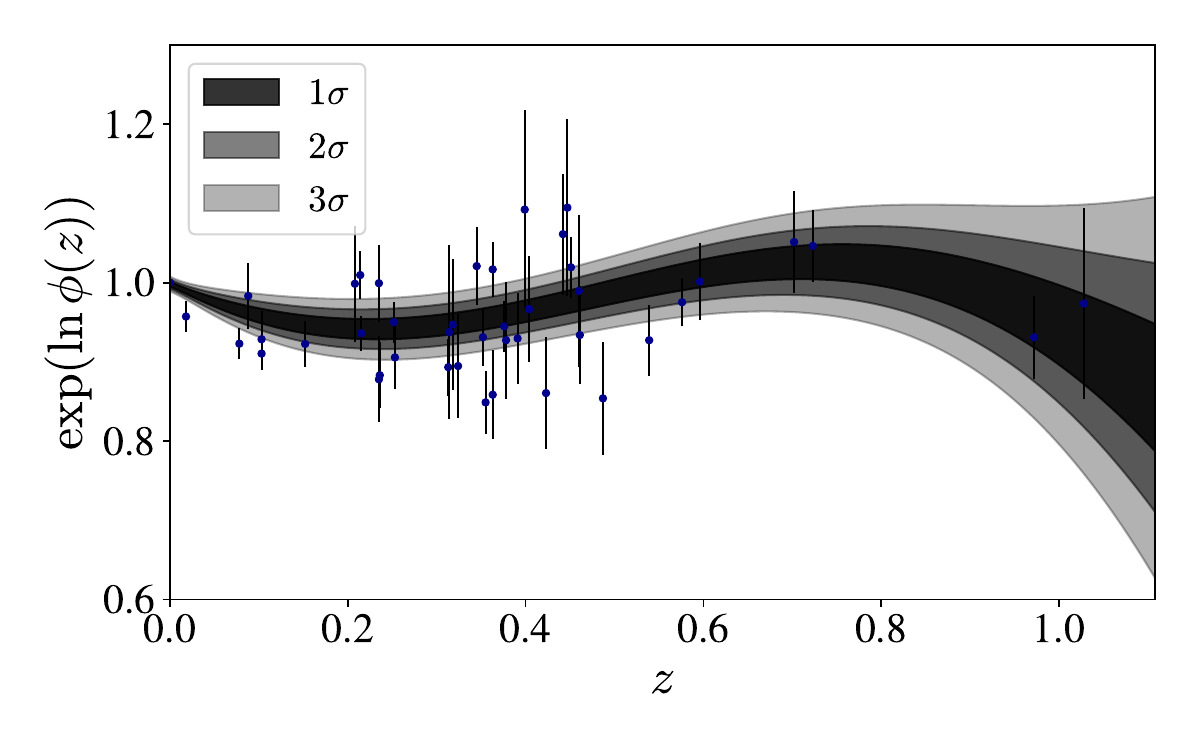}
\caption{Left Panel: the reconstruction of the function $\ln \phi(z)$. Right Panel: reconstruction of $\exp \left( \ln \phi(z) \right)$.}
\label{fig:result_ln}
\end{figure*}

It is well established that a detailed understanding of the thermodynamic state of the intracluster gas and the dark matter distribution is essential for using galaxy clusters as precise cosmological probes. In particular, cluster masses inferred under the hydrostatic equilibrium assumption are typically lower than those obtained from independent methods such as strong gravitational lensing (see \cite{Gonzalez2024qjs,Colaco2025yhw,Colaco2025aqp,Colaco2023gzy,Colaco2022noc,Colaco2020wbn,Colaco2020ndf}), which may lead to underestimated gas mass fractions and consequently to biased values of $\phi(z)$. In addition, constraints on a varying $G$ may be partially degenerate with other parameters, such as the absolute magnitude of SNe Ia ($M_B$), the Hubble constant ($H_0$). Nevertheless, since SNe Ia and cluster gas fractions depend on $G$ through different physical mechanisms, their joint analysis provides a complementary way to mitigate these degeneracies and to constrain possible variations of the gravitational constant.

\subsection{Robustness test using $\ln\phi(z)$ reconstruction}

{Since the gravitational coupling is written as $G(z)=G_0\phi(z)$, the function $\phi(z)$ is physically required to remain positive over the entire redshift range. Motivated by this property, we also reconstructed the quantity $\ln\phi(z)$ using Gaussian Processes as a robustness test. In this parametrization, the no-variation limit simply corresponds to $\ln\phi(z)=0$, while the positivity condition is naturally preserved in this parametrization. The left panel of Fig. \ref{fig:result_ln} shows the GP reconstruction of $\ln\phi(z)$ for the analyzed case $n=0.73$. Mild localized oscillatory features are present in the reconstructed mean function, particularly at intermediate redshifts. For visualization purposes, the exponential transformation was applied point-by-point to the reconstructed GP mean function and to the corresponding confidence regions. The resulting reconstruction is shown in the right panel of Fig.\ref{fig:result_ln}.

In order to evaluate the robustness of the reconstruction with respect to the choice of variable, we converted the reconstructed function back to $\phi(z)$ through the transformation
\[
\phi(z)=\exp[\ln\phi(z)].
\]

We then compared this indirect reconstruction of $\phi(z)$ with the direct GP reconstruction presented in the main analysis. We found that both approaches exhibit a qualitatively consistent behavior over the entire redshift range considered, despite the mild local oscillatory features present in the $\ln\phi(z)$ reconstruction. In particular, the reconstructed evolution of the gravitational coupling remains compatible with a constant value within the corresponding confidence intervals in both parametrizations. This agreement indicates that the inferred behavior of $G(z)$ is not significantly affected by the choice of reconstructed variable and supports the robustness of our results.

As previously discussed at the end of Section II, the standard scenario ($n=-3/4$) leads to a highly nonlinear dependence of $\phi(z)$ on the observational quantities, amplifying statistical fluctuations and resulting in unstable and weakly constrained reconstructions. We also applied the same $\ln\phi(z)$ reconstruction procedure to this case and obtained qualitatively the same statistical behavior, with excessively broad confidence regions and no statistically informative constraints on the evolution of $G(z)$.1
 }

\section{Conclusions}
\label{final}

In this paper, we explored a possible time evolution of the gravitational constant $G$ using a non-parametric approach. We combined measurements of the gas mass fraction of galaxy clusters with luminosity distances from SNe Ia, since both observables are sensitive to the value of the gravitational coupling. Variations in $G$ modify the Chandrasekhar mass of white dwarf progenitors, affecting the intrinsic luminosities of SNe Ia and the corresponding distance estimates. At the same time, the gas mass fraction in galaxy clusters depends on $G$ through its impact on cluster dynamics and total mass determinations. Consequently, the joint analysis of these probes provides an effective way to investigate a possible redshift evolution of $G$.

In particular, we considered a specific phenomenological scenario motivated by some scalar--tensor and screened modified-gravity frameworks, in which semi-analytic studies suggest that the standardized peak luminosity of SNe Ia decreases with increasing Chandrasekhar mass. Adopting the locally measured value of $M_B$, we obtained luminosity distance measurements from the Pantheon$+$ sample and reconstructed the function $\phi(z)$, defined through $G=G_0\phi(z)$, using Gaussian Processes.

Our results indicate that the reconstructed evolution of $G(z)$ remains compatible with a constant gravitational coupling over the redshift range analyzed. Although the reconstruction permits mild low-redshift departures from a constant $G$ ($z<0.4$), no statistically significant deviation is observed. These results highlight the potential of combining complementary astrophysical probes to investigate the stability of fundamental constants and possible varying-$G$ scenarios. Although current uncertainties remain considerable, future high-quality galaxy cluster observations from eROSITA are expected to substantially improve the constraints obtained in this work. In addition, future investigations exploring alternative cosmological observables combined with SNe Ia data to test the standard scenario ($n=-3/4$) are currently in progress and may help establish more robust constraints within the standard Chandrasekhar mass--luminosity framework.

\begin{acknowledgments}
\noindent RFLH thanks the financial support from the Conselho Nacional de Desenvolvimento Cientıfico e Tecnológico (CNPq) under the project No. 308550/2023-47.

\end{acknowledgments}

\bibliography{PRD}

@article{Dirac1937ti,
    author = "Dirac, Paul A. M.",
    title = "{The Cosmological constants}",
    doi = "10.1038/139323a0",
    journal = "Nature",
    volume = "139",
    pages = "323",
    year = "1937"
}

@ARTICLE{2025PhRvD.111f3526K,
       author = {{Krolewski}, Alex and {Percival}, Will J.},
        title = "{Measuring the baryon fraction using galaxy clustering}",
      journal = {\prd},
         year = 2025,
        month = mar,
       volume = {111},
       number = {6},
          eid = {063526},
        pages = {063526},
          doi = {10.1103/PhysRevD.111.063526},
archivePrefix = {arXiv},
       eprint = {2403.19236},
 primaryClass = {astro-ph.CO},
       adsurl = {https://ui.adsabs.harvard.edu/abs/2025PhRvD.111f3526K}
}

@article{Uzan2011,
author = {Uzan, Jean-Philippe},
title = {Varying Constants, Gravitation and Cosmology},
journal = {Living Reviews in Relativity},
volume = {14},
pages = {2},
year = {2011},
doi = {10.12942/lrr-2011-2}
}

@ARTICLE{2003RvMP...75..403U,
       author = {{Uzan}, Jean-Philippe},
        title = "{The fundamental constants and their variation: observational and theoretical status}",
      journal = {Reviews of Modern Physics},
         year = 2003,
        month = apr,
       volume = {75},
       number = {2},
        pages = {403-455},
          doi = {10.1103/RevModPhys.75.403},
archivePrefix = {arXiv},
       eprint = {hep-ph/0205340},
 primaryClass = {hep-ph},
       adsurl = {https://ui.adsabs.harvard.edu/abs/2003RvMP...75..403U}
}

@article{Uzan2002vq,
    author = "Uzan, Jean-Philippe",
    title = "{The Fundamental Constants and Their Variation: Observational Status and Theoretical Motivations}",
    eprint = "hep-ph/0205340",
    archivePrefix = "arXiv",
    doi = "10.1103/RevModPhys.75.403",
    journal = "Rev. Mod. Phys.",
    volume = "75",
    pages = "403",
    year = "2003"
}

@article{applegate2014weighing,
  title={Weighing the Giants--III. Methods and measurements of accurate galaxy cluster weak-lensing masses},
  author={Applegate, Douglas E and von der Linden, Anja and Kelly, Patrick L and Allen, Mark T and Allen, Steven W and Burchat, Patricia R and Burke, David L and Ebeling, Harald and Mantz, Adam and Morris, R Glenn},
  journal={Monthly Notices of the Royal Astronomical Society},
  volume={439},
  number={1},
  pages={48--72},
  year={2014},
  publisher={Oxford University Press}
}

@article{Holanda_2019,
   title={An estimate of the dark matter density from galaxy clusters and supernovae data},
   volume={2019},
   ISSN={1475-7516},
   url={http://dx.doi.org/10.1088/1475-7516/2019/11/032},
   DOI={10.1088/1475-7516/2019/11/032},
   number={11},
   journal={Journal of Cosmology and Astroparticle Physics},
   publisher={IOP Publishing},
   author={Holanda, R.F.L. and Gonçalves, R.S. and Gonzalez, J.E. and Alcaniz, J.S.},
   year={2019},
   month=nov, pages={032–032} }

@article{Uzan2010pm,
    author = "Uzan, Jean-Philippe",
    title = "{Varying Constants, Gravitation and Cosmology}",
    eprint = "1009.5514",
    archivePrefix = "arXiv",
    primaryClass = "astro-ph.CO",
    doi = "10.12942/lrr-2011-2",
    journal = "Living Rev. Rel.",
    volume = "14",
    pages = "2",
    year = "2011"
}

@article{Martins2017yxk,
    author = "Martins, C. J. A. P.",
    title = "{The status of varying constants: a review of the physics, searches and implications}",
    eprint = "1709.02923",
    archivePrefix = "arXiv",
    primaryClass = "astro-ph.CO",
    doi = "10.1088/1361-6633/aa860e",
    month = "9",
    year = "2017"
}

@article{Uzan2024ded,
    author = "Uzan, Jean-Philippe",
    title = "{Fundamental constants: from measurement to the universe, a window on gravitation and cosmology}",
    eprint = "2410.07281",
    archivePrefix = "arXiv",
    primaryClass = "astro-ph.CO",
    doi = "10.1007/s41114-025-00059-y",
    journal = "Living Rev. Rel.",
    volume = "28",
    number = "1",
    pages = "6",
    year = "2025"
}

@article{Gundlach2000rk,
    author = "Gundlach, Jens H. and Merkowitz, Stephen M.",
    title = "{Measurement of Newton's constant using a torsion balance with angular acceleration feedback}",
    eprint = "gr-qc/0006043",
    archivePrefix = "arXiv",
    doi = "10.1103/PhysRevLett.85.2869",
    journal = "Phys. Rev. Lett.",
    volume = "85",
    pages = "2869--2872",
    year = "2000"
}

@article{101093nsrnwaa165,
    author = {Xue, Chao and Liu, Jian-Ping and Li, Qing and Wu, Jun-Fei and Yang, Shan-Qing and Liu, Qi and Shao, Cheng-Gang and Tu, Liang-Cheng and Hu, Zhong-Kun and Luo, Jun},
    title = {Precision measurement of the Newtonian gravitational constant},
    journal = {National Science Review},
    volume = {7},
    number = {12},
    pages = {1803-1817},
    year = {2020},
    month = {07},
    issn = {2095-5138},
    doi = {10.1093/nsr/nwaa165},
    url = {https://doi.org/10.1093/nsr/nwaa165},
    eprint = {https://academic.oup.com/nsr/article-pdf/7/12/1803/38880653/nwaa165.pdf},
}

@article{Wang2012kj,
    author = "Wang, Junpu and Hui, Lam and Khoury, Justin",
    title = "{No-Go Theorems for Generalized Chameleon Field Theories}",
    eprint = "1208.4612",
    archivePrefix = "arXiv",
    primaryClass = "astro-ph.CO",
    doi = "10.1103/PhysRevLett.109.241301",
    journal = "Phys. Rev. Lett.",
    volume = "109",
    pages = "241301",
    year = "2012"
}

@article{PhysRevLett852869,
  title = {Measurement of Newton's Constant Using a Torsion Balance with Angular Acceleration Feedback},
  author = {Gundlach, Jens H. and Merkowitz, Stephen M.},
  journal = {Phys. Rev. Lett.},
  volume = {85},
  issue = {14},
  pages = {2869--2872},
  numpages = {0},
  year = {2000},
  month = {Oct},
  publisher = {American Physical Society},
  doi = {10.1103/PhysRevLett.85.2869},
  url = {https://link.aps.org/doi/10.1103/PhysRevLett.85.2869}
}

@article{Tino2020dsl,
    author = "Tino, Guglielmo M.",
    title = "{Testing gravity with cold atom interferometry: Results and prospects}",
    eprint = "2009.01484",
    archivePrefix = "arXiv",
    primaryClass = "gr-qc",
    doi = "10.1088/2058-9565/abd83e",
    journal = "Quantum Sci. Technol.",
    volume = "6",
    number = "2",
    pages = "024014",
    year = "2021"
}

@article{Umilta2015cta,
    author = "Umilt{\`a}, C. and Ballardini, M. and Finelli, F. and Paoletti, D.",
    title = "{CMB and BAO constraints for an induced gravity dark energy model with a quartic potential}",
    eprint = "1507.00718",
    archivePrefix = "arXiv",
    primaryClass = "astro-ph.CO",
    doi = "10.1088/1475-7516/2015/08/017",
    journal = "JCAP",
    volume = "08",
    pages = "017",
    year = "2015"
}

@article{Ballardini2016cvy,
    author = "Ballardini, Mario and Finelli, Fabio and Umilt{\`a}, Caterina and Paoletti, Daniela",
    title = "{Cosmological constraints on induced gravity dark energy models}",
    eprint = "1601.03387",
    archivePrefix = "arXiv",
    primaryClass = "astro-ph.CO",
    doi = "10.1088/1475-7516/2016/05/067",
    journal = "JCAP",
    volume = "05",
    pages = "067",
    year = "2016"
}

@article{Xue2014kna,
    author = "Xue, She-Sheng",
    title = "{How universe evolves with cosmological and gravitational constants}",
    eprint = "1410.6152",
    archivePrefix = "arXiv",
    primaryClass = "gr-qc",
    doi = "10.1016/j.nuclphysb.2015.05.022",
    journal = "Nucl. Phys. B",
    volume = "897",
    pages = "326--345",
    year = "2015"
}

@article{Gelmini2020ekg,
    author = "Gelmini, Graciela B. and Kawasaki, Masahiro and Kusenko, Alexander and Murai, Kai and Takhistov, Volodymyr",
    title = "{Big Bang Nucleosynthesis constraints on sterile neutrino and lepton asymmetry of the Universe}",
    eprint = "2005.06721",
    archivePrefix = "arXiv",
    primaryClass = "hep-ph",
    reportNumber = "IPMU 20-0051",
    doi = "10.1088/1475-7516/2020/09/051",
    journal = "JCAP",
    volume = "09",
    pages = "051",
    year = "2020"
}

@article{Zhang2017aqn,
    author = "Zhang, Bonnie R. and Childress, Michael J. and Davis, Tamara M. and Karpenka, Natallia V. and Lidman, Chris and Schmidt, Brian P. and Smith, Mathew",
    title = "{A blinded determination of $H_0$ from low-redshift Type Ia supernovae, calibrated by Cepheid variables}",
    eprint = "1706.07573",
    archivePrefix = "arXiv",
    primaryClass = "astro-ph.CO",
    doi = "10.1093/mnras/stx1600",
    journal = "Mon. Not. Roy. Astron. Soc.",
    volume = "471",
    number = "2",
    pages = "2254--2285",
    year = "2017"
}

@article{Wright2017rsu,
    author = "Wright, Bill S. and Li, Baojiu",
    title = "{Type Ia supernovae, standardizable candles, and gravity}",
    eprint = "1710.07018",
    archivePrefix = "arXiv",
    primaryClass = "astro-ph.CO",
    doi = "10.1103/PhysRevD.97.083505",
    journal = "Phys. Rev. D",
    volume = "97",
    number = "8",
    pages = "083505",
    year = "2018"
}

@article{Zhao2018gwk,
    author = "Zhao, Wen and Wright, Bill S. and Li, Baojiu",
    title = "{Constraining the time variation of Newton's constant $G$ with gravitational-wave standard sirens and supernovae}",
    eprint = "1804.03066",
    archivePrefix = "arXiv",
    primaryClass = "astro-ph.CO",
    doi = "10.1088/1475-7516/2018/10/052",
    journal = "JCAP",
    volume = "10",
    pages = "052",
    year = "2018"
}

@article{Vijaykumar2020nzc,
    author = "Vijaykumar, Aditya and Kapadia, Shasvath J. and Ajith, Parameswaran",
    title = "{Constraints on the time variation of the gravitational constant using gravitational-wave observations of binary neutron stars}",
    eprint = "2003.12832",
    archivePrefix = "arXiv",
    primaryClass = "gr-qc",
    doi = "10.1103/PhysRevLett.126.141104",
    journal = "Phys. Rev. Lett.",
    volume = "126",
    number = "14",
    pages = "141104",
    year = "2021"
}

@ARTICLE{2011ARAA49409A,
       author = {{Allen}, Steven W. and {Evrard}, August E. and {Mantz}, Adam B.},
        title = "{Cosmological Parameters from Observations of Galaxy Clusters}",
      journal = {ARAA},
         year = 2011,
        month = sep,
       volume = {49},
       number = {1},
        pages = {409-470},
          doi = {10.1146/annurev-astro-081710-102514},
archivePrefix = {arXiv},
       eprint = {1103.4829},
 primaryClass = {astro-ph.CO},
       adsurl = {https://ui.adsabs.harvard.edu/abs/2011ARA&A..49..409A}
}

@article{Amendola1999vu,
    author = "Amendola, Luca and Corasaniti, Pier Stefano and Occhionero, Franco",
    title = "{Time variability of the gravitational constant and type Ia supernovae}",
    eprint = "astro-ph/9907222",
    archivePrefix = "arXiv",
    month = "7",
    year = "1999"
}

@article{Alvey2019ctk,
    author = "Alvey, James and Sabti, Nashwan and Escudero, Miguel and Fairbairn, Malcolm",
    title = "{Improved BBN Constraints on the Variation of the Gravitational Constant}",
    eprint = "1910.10730",
    archivePrefix = "arXiv",
    primaryClass = "astro-ph.CO",
    doi = "10.1140/epjc/s10052-020-7727-y",
    journal = "Eur. Phys. J. C",
    volume = "80",
    number = "2",
    pages = "148",
    year = "2020"
}

@article{Ferreira2024kht,
    author = "Ferreira, Marcelo and Holanda, Rodrigo F. L. and Gonzalez, Javier E. and Cola{\c{c}}o, L. R. and Nunes, Rafael C.",
    title = "{Non-parametric reconstruction of the fine structure constant with galaxy clusters}",
    eprint = "2410.21542",
    archivePrefix = "arXiv",
    primaryClass = "astro-ph.CO",
    reportNumber = "84,1120 (2024) EPJC",
    doi = "10.1140/epjc/s10052-024-13468-0",
    journal = "Eur. Phys. J. C",
    volume = "84",
    number = "10",
    pages = "1120",
    year = "2024"
}

@article{Colaco2020ndf,
    author = "Cola{\c{c}}o, L. R. and Holanda, R. F. L. and Silva, R.",
    title = "{Probing variation of the fine-structure constant in runaway dilaton models using Strong Gravitational Lensing and Type Ia Supernovae}",
    eprint = "2004.08484",
    archivePrefix = "arXiv",
    primaryClass = "astro-ph.CO",
    doi = "10.1140/epjc/s10052-021-09625-4",
    journal = "Eur. Phys. J. C",
    volume = "81",
    number = "9",
    pages = "822",
    year = "2021"
}

@article{Colaco2020wbn,
    author = "Cola{\c{c}}o, L. R. and Gonzalez, J. E. and Holanda, R. F. L.",
    title = "{Gravitational lens time-delay as a probe of a possible time variation of the fine-structure constant}",
    eprint = "2010.04021",
    archivePrefix = "arXiv",
    primaryClass = "astro-ph.CO",
    doi = "10.1140/epjc/s10052-021-09319-x",
    journal = "Eur. Phys. J. C",
    volume = "81",
    number = "6",
    pages = "533",
    year = "2021"
}

@article{Brax2008hh,
    author = "Brax, Philippe and van de Bruck, Carsten and Davis, Anne-Christine and Shaw, Douglas J.",
    title = "{f(R) Gravity and Chameleon Theories}",
    eprint = "0806.3415",
    archivePrefix = "arXiv",
    primaryClass = "astro-ph",
    doi = "10.1103/PhysRevD.78.104021",
    journal = "Phys. Rev. D",
    volume = "78",
    pages = "104021",
    year = "2008"
}

@article{Mould2014iga,
    author = "Mould, Jeremy and Uddin, Syed",
    title = "{Constraining a possible variation of G with Type Ia supernovae}",
    eprint = "1402.1534",
    archivePrefix = "arXiv",
    primaryClass = "astro-ph.CO",
    doi = "10.1017/pasa.2014.9",
    journal = "Publ. Astron. Soc. Austral.",
    volume = "31",
    pages = "15",
    year = "2014"
}

@article{LorenAguilar2003qtx,
    author = "Loren-Aguilar, P. and Garcia-Berro, E. and Isern, J. and Kubyshin, Yu. A.",
    title = "{Time variation of G and alpha within models with extra dimensions}",
    eprint = "astro-ph/0309722",
    archivePrefix = "arXiv",
    doi = "10.1088/0264-9381/20/18/302",
    journal = "Class. Quant. Grav.",
    volume = "20",
    pages = "3885--3896",
    year = "2003"
}

@article{Colaco2023gzy,
    author = "Cola{\c{c}}o, L. R. and Ferreira, Marcelo and Holanda, R. F. L. and Gonzalez, J. E. and Nunes, Rafael C.",
    title = "{A Hubble constant estimate from galaxy cluster and type Ia SNe observations}",
    eprint = "2310.18711",
    archivePrefix = "arXiv",
    primaryClass = "astro-ph.CO",
    doi = "10.1088/1475-7516/2024/05/098",
    journal = "JCAP",
    volume = "05",
    pages = "098",
    year = "2024"
}

@article{Colaco2025aqp,
    author = "Cola{\c{c}}o, Leonardo R.",
    title = "{A Hubble Constant Determination Through Quasar Time Delays and Type Ia Supernovae}",
    eprint = "2503.06189",
    archivePrefix = "arXiv",
    primaryClass = "astro-ph.CO",
    doi = "10.3390/universe11030089",
    journal = "Universe",
    volume = "11",
    number = "3",
    pages = "89",
    year = "2025"
}

@article{Colaco2025yhw,
    author = "Cola{\c{c}}o, L. R. and Holanda, R. F. L. and Santana, Z. C. and Silva, R.",
    title = "{A joint analysis of strong lensing and type Ia supernovae to determine the Hubble constant}",
    eprint = "2505.17262",
    archivePrefix = "arXiv",
    primaryClass = "astro-ph.CO",
    doi = "10.1140/epjc/s10052-025-14315-6",
    journal = "Eur. Phys. J. C",
    volume = "85",
    number = "5",
    pages = "577",
    year = "2025"
}

@article{Gonzalez2024qjs,
    author = "Gonzalez, Javier E. and Ferreira, Marcelo and Cola{\c{c}}o, Leorando R. and Holanda, Rodrigo F. L. and Nunes, Rafael C.",
    title = "{Unveiling the Hubble constant through galaxy cluster gas mass fractions}",
    eprint = "2405.13665",
    archivePrefix = "arXiv",
    primaryClass = "astro-ph.CO",
    doi = "10.1016/j.physletb.2024.138982",
    journal = "Phys. Lett. B",
    volume = "857",
    pages = "138982",
    year = "2024"
}

@article{101093mnrasstab3390,
    author = {Mantz, Adam B and Morris, R Glenn and Allen, Steven W and Canning, Rebecca E A and Baumont, Lucie and Benson, Bradford and Bleem, Lindsey E and Ehlert, Steven R and Floyd, Benjamin and Herbonnet, Ricardo and Kelly, Patrick L and Liang, Shuang and von der Linden, Anja and McDonald, Michael and Rapetti, David A and Schmidt, Robert W and Werner, Norbert and Wright, Adam},
    title = {Cosmological constraints from gas mass fractions of massive, relaxed galaxy clusters},
    journal = {Monthly Notices of the Royal Astronomical Society},
    volume = {510},
    number = {1},
    pages = {131-145},
    year = {2021},
    month = {11},
    issn = {0035-8711},
    doi = {10.1093/mnras/stab3390},
    url = {https://doi.org/10.1093/mnras/stab3390},
    eprint = {https://academic.oup.com/mnras/article-pdf/510/1/131/41796169/stab3390.pdf},
}

@article{Scolnic2021amr,
    author = "Scolnic, Dan and others",
    title = "{The Pantheon+ Analysis: The Full Data Set and Light-curve Release}",
    eprint = "2112.03863",
    archivePrefix = "arXiv",
    primaryClass = "astro-ph.CO",
    doi = "10.3847/1538-4357/ac8b7a",
    journal = "Astrophys. J.",
    volume = "938",
    number = "2",
    pages = "113",
    year = "2022"
}

@article{Ruchika2024ymt,
    author = "Ruchika and Perivolaropoulos, Leandros and Melchiorri, Alessandro",
    title = "{Effects of a local physics change on the SH0ES determination of H0}",
    eprint = "2408.03875",
    archivePrefix = "arXiv",
    primaryClass = "astro-ph.CO",
    doi = "10.1103/19pn-3bvs",
    journal = "Phys. Rev. D",
    volume = "111",
    number = "12",
    pages = "123526",
    year = "2025"
}

@ARTICLE{2026PhRvD.113b3501P,
       author = {{Perivolaropoulos}, Leandros and {Ruchika}},
        title = "{Hubble tension and the G-step model: Reexamination of recent constraints on modified local physics}",
      journal = {\prd},
         year = 2026,
        month = jan,
       volume = {113},
       number = {2},
          eid = {023501},
        pages = {023501},
          doi = {10.1103/mgyd-6cf2},
       adsurl = {https://ui.adsabs.harvard.edu/abs/2026PhRvD.113b3501P}
}

@article{Khoury2003rn,
    author = "Khoury, Justin and Weltman, Amanda",
    title = "{Chameleon cosmology}",
    eprint = "astro-ph/0309411",
    archivePrefix = "arXiv",
    doi = "10.1103/PhysRevD.69.044026",
    journal = "Phys. Rev. D",
    volume = "69",
    pages = "044026",
    year = "2004"
}

@article{Khoury2003aq,
    author = "Khoury, Justin and Weltman, Amanda",
    title = "{Chameleon fields: Awaiting surprises for tests of gravity in space}",
    eprint = "astro-ph/0309300",
    archivePrefix = "arXiv",
    doi = "10.1103/PhysRevLett.93.171104",
    journal = "Phys. Rev. Lett.",
    volume = "93",
    pages = "171104",
    year = "2004"
}

@article{CeronHurtado2016jrp,
    author = "Ceron-Hurtado, Juan J. and He, Jian-hua and Li, Baojiu",
    title = "{Can background cosmology hold the key for modified gravity tests?}",
    eprint = "1609.00532",
    archivePrefix = "arXiv",
    primaryClass = "astro-ph.CO",
    doi = "10.1103/PhysRevD.94.064052",
    journal = "Phys. Rev. D",
    volume = "94",
    number = "6",
    pages = "064052",
    year = "2016"
}

@article{Ruchika2023ugh,
    author = "Ruchika and Rathore, Himansh and Roy Choudhury, Shouvik and Rentala, Vikram",
    title = "{A gravitational constant transition within cepheids as supernovae calibrators can solve the Hubble tension}",
    eprint = "2306.05450",
    archivePrefix = "arXiv",
    primaryClass = "astro-ph.CO",
    doi = "10.1088/1475-7516/2024/06/056",
    journal = "JCAP",
    volume = "06",
    pages = "056",
    year = "2024"
}

@article{Lamine2024xno,
    author = "Lamine, Brahim and Ozdalkiran, Yacob and Mirouze, Louis and Erdogan, Furkan and Ilic, St{\'e}phane and Tutusaus, Isaac and Kou, Raphael and Blanchard, Alain",
    title = "{Cosmological measurement of the gravitational constant G using the CMB, BAO, and BBN}",
    eprint = "2407.15553",
    archivePrefix = "arXiv",
    primaryClass = "astro-ph.CO",
    doi = "10.1051/0004-6361/202451602",
    journal = "Astron. Astrophys.",
    volume = "697",
    pages = "A109",
    year = "2025"
}

@article{Holanda2025xsj,
    author = "Holanda, R. F. L. and Ferreira, Marcelo and Gonzalez, Javier E.",
    title = "{An investigation of a varying G through Strong Lensing and SNe Ia observations}",
    eprint = "2508.00075",
    archivePrefix = "arXiv",
    primaryClass = "astro-ph.CO",
    doi = "10.1016/j.physletb.2025.139756",
    journal = "Phys. Lett. B",
    volume = "868",
    pages = "139756",
    year = "2025"
}

@article{BezerraSobrinho2025vaf,
    author = "Bezerra-Sobrinho, J. and Cuzinatto, R. R. and Medeiros, L. G. and Pompeia, P. J.",
    title = "{Brans-Dicke-like field for co-varying $G$ and $c$: observational constraints}",
    eprint = "2509.09211",
    archivePrefix = "arXiv",
    primaryClass = "astro-ph.CO",
    month = "9",
    year = "2025"
}

@ARTICLE{2019PhRvD.100j4035S,
       author = {{Sakstein}, Jeremy and {Desmond}, Harry and {Jain}, Bhuvnesh},
        title = "{Screened fifth forces mediated by dark matter-baryon interactions: Theory and astrophysical probes}",
      journal = {\prd},
         year = 2019,
        month = nov,
       volume = {100},
       number = {10},
          eid = {104035},
        pages = {104035},
          doi = {10.1103/PhysRevD.100.104035},
archivePrefix = {arXiv},
       eprint = {1907.03775},
 primaryClass = {astro-ph.CO},
       adsurl = {https://ui.adsabs.harvard.edu/abs/2019PhRvD.100j4035S}
}

@article{Brans:1961sx,
    author = "Brans, C. and Dicke, R. H.",
    editor = "Hsu, Jong-Ping and Fine, D.",
    title = "{Mach's principle and a relativistic theory of gravitation}",
    doi = "10.1103/PhysRev.124.925",
    journal = "Phys. Rev.",
    volume = "124",
    pages = "925--935",
    year = "1961"
}

@article{PhysRevD66023525,
  title = {Cosmological observations in scalar-tensor quintessence},
  author = {Riazuelo, Alain and Uzan, Jean-Philippe},
  journal = {Phys. Rev. D},
  volume = {66},
  issue = {2},
  pages = {023525},
  numpages = {18},
  year = {2002},
  month = {Jul},
  publisher = {American Physical Society},
  doi = {10.1103/PhysRevD.66.023525},
  url = {https://link.aps.org/doi/10.1103/PhysRevD.66.023525}
}

@article{Bai2015vca,
    author = "Bai, Yang and Salvado, Jordi and Stefanek, Ben A.",
    title = "{Cosmological Constraints on the Gravitational Interactions of Matter and Dark Matter}",
    eprint = "1505.04789",
    archivePrefix = "arXiv",
    primaryClass = "hep-ph",
    doi = "10.1088/1475-7516/2015/10/029",
    journal = "JCAP",
    volume = "10",
    pages = "029",
    year = "2015"
}

@ARTICLE{Ettori2009,
       author = {{Ettori}, S. and {Morandi}, A. and {Tozzi}, P. and {Balestra}, I. and {Borgani}, S. and {Rosati}, P. and {Lovisari}, L. and {Terenziani}, F.},
        title = "{The cluster gas mass fraction as a cosmological probe: a revised study}",
      journal = {aap},
         year = 2009,
        month = jul,
       volume = {501},
       number = {1},
        pages = {61-73},
          doi = {10.1051/0004-6361/200810878},
archivePrefix = {arXiv},
       eprint = {0904.2740},
 primaryClass = {astro-ph.CO},
       adsurl = {https://ui.adsabs.harvard.edu/abs/2009A&A...501...61E}
}

@BOOK{sarazin,
       author = {{Sarazin}, Craig L.},
        title = "{X-ray emission from clusters of galaxies}",
         year = 1988,
       adsurl = {https://ui.adsabs.harvard.edu/abs/1988xrec.book.....S}
}

@article{Seikel,
      author         = "Seikel, Marina and Clarkson, Chris and Smith, Mathew",
      title          = "{Reconstruction of dark energy and expansion dynamics
                        using Gaussian processes}",
      journal        = "JCAP",
      volume         = "1206",
      year           = "2012",
      pages          = "036",
      doi            = "10.1088/1475-7516/2012/06/036",
      eprint         = "1204.2832",
      archivePrefix  = "arXiv",
      primaryClass   = "astro-ph.CO",
      SLACcitation   = "%%CITATION = ARXIV:1204.2832;%%"
}

@article{Holanda2017,
author = {Holanda, R. F. L. and Busti, V. C. and Gonzalez, J. E. and Andrade-Santos, F. and Alcaniz, J. S.},
title = {Cosmological constraints on the gas depletion factor in galaxy clusters},
year = {2017},
eprint = {1706.07321},
archivePrefix = {arXiv},
primaryClass = {astro-ph.CO}
}

@ARTICLE{2013ApJ777123B,
       author = {Battaglia, N. and Bond, J.R. and Pfrommer, C. and Sievers, J.L.},
        title = "{On the Cluster Physics of Sunyaev-Zel'dovich and X-Ray Surveys. III. Measurement Biases and Cosmological Evolution of Gas and Stellar Mass Fractions}",
      journal = {apj},
         year = {2013},
        month = {nov},
       volume = {777},
       number = {2},
          eid = {123},
        pages = {123},
          doi = {10.1088/0004-637X/777/2/123},
archivePrefix = {arXiv},
       eprint = {1209.4082},
 primaryClass = {astro-ph.CO},
       adsurl = {https://ui.adsabs.harvard.edu/abs/2013ApJ...777..123B}
}

@article{Mantz:2014xba,
    author = "Mantz, A. B. and Allen, S. W. and Morris, R. G. and Rapetti, D. A. and Applegate, D. E. and Kelly, P. L. and von der Linden, Anja and Schmidt, R. W.",
    title = "{Cosmology and astrophysics from relaxed galaxy clusters \textendash{} II. Cosmological constraints}",
    eprint = "1402.6212",
    archivePrefix = "arXiv",
    primaryClass = "astro-ph.CO",
    doi = "10.1093/mnras/stu368",
    journal = "Mon. Not. Roy. Astron. Soc.",
    volume = "440",
    number = "3",
    pages = "2077--2098",
    year = "2014"
}

@article{woosley1986physics,
  title={The physics of supernova explosions},
  author={Woosley, S Eo and Weaver, Thomas A},
  journal={IN: Annual review of astronomy and astrophysics. Volume 24 (A87-26730 10-90). Palo Alto, CA, Annual Reviews, Inc., 1986, p. 205-253.},
  volume={24},
  pages={205--253},
  year={1986}
}

@article{gaztanaga2001bounds,
  title={Bounds on the possible evolution of the gravitational constant from cosmological type-Ia supernovae},
  author={Gaztanaga, Enrique and Garcia-Berro, Enrique and Isern, Jordi and Bravo, Eduardo and Dominguez, Inma},
  journal={Physical Review D},
  volume={65},
  number={2},
  pages={023506},
  year={2001},
  publisher={APS}
}

@article{riess2022comprehensive,
  title={A comprehensive measurement of the local value of the Hubble constant with 1 km s- 1 Mpc- 1 uncertainty from the Hubble Space Telescope and the SH0ES team},
  author={Riess, Adam G and Yuan, Wenlong and Macri, Lucas M and Scolnic, Dan and Brout, Dillon and Casertano, Stefano and Jones, David O and Murakami, Yukei and Anand, Gagandeep S and Breuval, Louise and others},
  journal={The Astrophysical journal letters},
  volume={934},
  number={1},
  pages={L7},
  year={2022},
  publisher={IOP Publishing}
}

@article{colacco2022varying,
  title={Varying-$\alpha$ in scalar--tensor theory: implications in light of the supernova absolute magnitude tension and forecast from GW standard sirens},
  author={Cola{\c{c}}o, LR and Holanda, RFL and Nunes, Rafael C},
  journal={arXiv preprint arXiv:2201.04073},
  year={2022}
}

@article{planelles2013baryon,
  title={Baryon census in hydrodynamical simulations of galaxy clusters},
  author={Planelles, Susana and Borgani, Stefano and Dolag, Klaus and Ettori, Stefano and Fabjan, Dunja and Murante, Giuseppe and Tornatore, Luca},
  journal={Monthly Notices of the Royal Astronomical Society},
  volume={431},
  number={2},
  pages={1487--1502},
  year={2013},
  publisher={Oxford University Press}
}

@article{allen2008improved,
  title={Improved constraints on dark energy from Chandra X-ray observations of the largest relaxed galaxy clusters},
  author={Allen, SW and Rapetti, DA and Schmidt, RW and Ebeling, H and Morris, RG and Fabian, AC},
  journal={Monthly Notices of the Royal Astronomical Society},
  volume={383},
  number={3},
  pages={879--896},
  year={2008},
  publisher={Blackwell Publishing Ltd Oxford, UK}
}

@article{Gupta2021tma,
    author = "Gupta, Rajendra P.",
    title = "{Effect of evolving physical constants on type Ia supernova luminosity}",
    eprint = "2112.10654",
    archivePrefix = "arXiv",
    primaryClass = "gr-qc",
    doi = "10.1093/mnras/stac254",
    journal = "Mon. Not. Roy. Astron. Soc.",
    volume = "511",
    number = "3",
    pages = "4238--4250",
    year = "2022"
}

@article{Ballardini2021eox,
    author = "Ballardini, Mario and Finelli, Fabio",
    title = "{Type Ia supernovae data with scalar-tensor gravity}",
    eprint = "2112.15126",
    archivePrefix = "arXiv",
    primaryClass = "astro-ph.CO",
    doi = "10.1103/PhysRevD.106.063531",
    journal = "Phys. Rev. D",
    volume = "106",
    number = "6",
    pages = "063531",
    year = "2022"
}

@article{GarciaBerro1999cwy,
    author = "Garcia-Berro, E. and Gaztanaga, E. and Isern, J. and Benvenuto, O. and Althaus, L.",
    title = "{On the evolution of cosmological type ia supernovae and the gravitational constant}",
    eprint = "astro-ph/9907440",
    archivePrefix = "arXiv",
    reportNumber = "IEEC-99-25",
    month = "7",
    year = "1999"
}

@article{Li2015aug,
    author = "Li, Ji-Xia and Wu, Feng-Quan and Li, Yi-Chao and Gong, Yan and Chen, Xue-Lei",
    title = "{Cosmological constraint on Brans-Dicke Model}",
    eprint = "1511.05280",
    archivePrefix = "arXiv",
    primaryClass = "astro-ph.CO",
    doi = "10.1088/1674-4527/15/12/003",
    journal = "Res. Astron. Astrophys.",
    volume = "15",
    number = "12",
    pages = "2151--2163",
    year = "2015"
}

@article{Hamity2005ri,
    author = "Hamity, Victor H. and Barraco, Daniel E.",
    title = "{A Stellar explosion in the weak field approximation of the Brans-Dicke theory}",
    eprint = "gr-qc/0504105",
    archivePrefix = "arXiv",
    doi = "10.1088/0264-9381/22/19/003",
    journal = "Class. Quant. Grav.",
    volume = "22",
    pages = "3841--3852",
    year = "2005"
}

@article{Sakstein2019qgn,
    author = "Sakstein, Jeremy and Desmond, Harry and Jain, Bhuvnesh",
    title = "{Screened Fifth Forces Mediated by Dark Matter--Baryon Interactions: Theory and Astrophysical Probes}",
    eprint = "1907.03775",
    archivePrefix = "arXiv",
    primaryClass = "astro-ph.CO",
    doi = "10.1103/PhysRevD.100.104035",
    journal = "Phys. Rev. D",
    volume = "100",
    number = "10",
    pages = "104035",
    year = "2019"
}

@article{Mantz2014xba,
    author = "Mantz, A. B. and Allen, S. W. and Morris, R. G. and Rapetti, D. A. and Applegate, D. E. and Kelly, P. L. and von der Linden, Anja and Schmidt, R. W.",
    title = "{Cosmology and astrophysics from relaxed galaxy clusters \textendash{} II. Cosmological constraints}",
    eprint = "1402.6212",
    archivePrefix = "arXiv",
    primaryClass = "astro-ph.CO",
    doi = "10.1093/mnras/stu368",
    journal = "Mon. Not. Roy. Astron. Soc.",
    volume = "440",
    number = "3",
    pages = "2077--2098",
    year = "2014"
}

@article{battaglia2013,
  title={ON THE CLUSTER PHYSICS OF SUNYAEV--ZEL'DOVICH AND X-RAY SURVEYS. III. MEASUREMENT BIASES AND COSMOLOGICAL EVOLUTION OF GAS AND STELLAR MASS FRACTIONS},
  author={Battaglia, N and Bond, JR and Pfrommer, C and Sievers, JL},
  journal={The Astrophysical Journal},
  volume={777},
  number={2},
  pages={123},
  year={2013},
  publisher={IOP Publishing}
}

@article{Seikel:2012uu,
    author = "Seikel, Marina and Clarkson, Chris and Smith, Mathew",
    title = "{Reconstruction of dark energy and expansion dynamics using Gaussian processes}",
    eprint = "1204.2832",
    archivePrefix = "arXiv",
    primaryClass = "astro-ph.CO",
    doi = "10.1088/1475-7516/2012/06/036",
    journal = "JCAP",
    volume = "06",
    pages = "036",
    year = "2012"
}

@ARTICLE{1996PASJ48L119S,
       author = {{Sasaki}, Shin},
        title = "{A New Method to Estimate Cosmological Parameters Using the Baryon Fraction of Clusters of Galaxies}",
      journal = {PASJ},
         year = 1996,
        month = dec,
       volume = {48},
        pages = {L119-L122},
          doi = {10.1093/pasj/48.6.L119},
archivePrefix = {arXiv},
       eprint = {astro-ph/9611033},
 primaryClass = {astro-ph},
       adsurl = {https://ui.adsabs.harvard.edu/abs/1996PASJ...48L.119S}
}

@ARTICLE{2017PhRvD95h4006H,
       author = {{Holanda}, R.~F.~L. and {Pereira}, S.~H. and {Santos da Costa}, S.},
        title = "{Searching for deviations from the general relativity theory with gas mass fraction of galaxy clusters and complementary probes}",
      journal = {prd},
         year = 2017,
        month = apr,
       volume = {95},
       number = {8},
          eid = {084006},
        pages = {084006},
          doi = {10.1103/PhysRevD.95.084006},
archivePrefix = {arXiv},
       eprint = {1612.09365},
 primaryClass = {astro-ph.CO}
}

@article{Colaco2022noc,
    author = "Cola\c{c}o, L. R. and Landau, S. J. and Gonzalez, J. E. and Spinelly, J. and Santos, G. L. F.",
    title = "{Constraining a possible time-variation of the speed of light along with the fine-structure constant using strong gravitational lensing and Type Ia supernovae observations}",
    eprint = "2204.06459",
    archivePrefix = "arXiv",
    primaryClass = "astro-ph.CO",
    doi = "10.1088/1475-7516/2022/08/062",
    journal = "JCAP",
    volume = "08",
    number = "08",
    pages = "062",
    year = "2022"
}

\end{document}